\documentclass[11pt]{article}

\usepackage[preprint]{acl}

\usepackage{times}
\usepackage{latexsym}

\usepackage[T1]{fontenc}

\usepackage[utf8]{inputenc}
\usepackage{newunicodechar}
\newunicodechar{‑}{\mbox{-}}
\newunicodechar{–}{\textendash}
\newunicodechar{’}{\textquoteright}
\newunicodechar{§}{\S}
\newunicodechar{é}{\'{e}}

\usepackage{microtype}

\usepackage{inconsolata}

\usepackage{booktabs}

\usepackage{graphicx}
\usepackage{enumitem} 
\usepackage[english]{babel}
\usepackage{blindtext}
\usepackage{algorithmic}
\usepackage{graphicx}
\usepackage{amsmath}
\usepackage{subcaption}
\usepackage[ruled, linesnumbered]{algorithm2e}
\usepackage{multirow}
\usepackage{hhline}
\usepackage{makecell}

\newcommand{\para}[1]{\noindent {\bf #1}}

\AtBeginDocument{%
  }

\def\name{\textsc{SpecMoE}\xspace}

\newcommand{\clk}[1]{{\color{black}  #1}}

\newcommand{\BULLET}{\vspace{+.00in} \noindent $\bullet$ \hspace{+.00in}}

\title{\name: Accelerating MoE Model Inference via Speculative Decoding and Pipelined Prefetching}

\author{
  Liangkun Chen\textsuperscript{1}, Zijian Wen\textsuperscript{1},
  Tian Wu\textsuperscript{1}, Chuan Wu\textsuperscript{2}, \\
  \bfseries Ennan Zhai\textsuperscript{3}, Xu Chen\textsuperscript{1},
  Xiaoxi Zhang\textsuperscript{1}\thanks{Corresponding author: Xiaoxi Zhang.} \\[0.4em]
  \textsuperscript{1}Sun Yat-sen University, Guangzhou, China \\
  \textsuperscript{2}The University of Hong Kong, Hong Kong, China \\
  \textsuperscript{3}Alibaba Cloud, Hangzhou, China \\[0.4em]
  {\normalfont\small\texttt{chenlk8@mail2.sysu.edu.cn}\quad\texttt{wenzj7@mail2.sysu.edu.cn}} \\
  {\normalfont\small\texttt{wutian@mail2.sysu.edu.cn}\quad\texttt{cwu@cs.hku.hk}\quad\texttt{ennan.zhai@alibaba-inc.com}} \\
  {\normalfont\small\texttt{chenxu35@mail.sysu.edu.cn}\quad\texttt{zhangxx89@mail.sysu.edu.cn}}
}

\hypersetup{
  pdftitle={SpecMoE: Accelerating MoE Model Inference via Speculative Decoding and Pipelined Prefetching},
  pdfauthor={Liangkun Chen, Zijian Wen, Tian Wu, Chuan Wu, Ennan Zhai, Xu Chen, Xiaoxi Zhang}
}

\begin{document}
\maketitle
\begin{abstract}
The Mixture-of-Experts (MoE) architecture has been widely adopted in large language models (LLMs) to reduce computation cost through model sparsity. Employing speculative decoding (SD) can further accelerate MoE inference by drafting multiple tokens per step and verifying them in parallel. However, combining MoE with SD inflates GPU memory and creates CPU–GPU bandwidth bottleneck during multi‑token verification. Existing MoE offloading systems are 
SD‑agnostic and do not address this bottleneck. We present \name, a novel SD-aware expert‑offloading framework with prefetching and compute–communication pipelining. \name introduces: (1) speculative expert prefetching that exploits structural correspondence between the draft and target models to prefetch likely experts ahead of verification; (2) a cutoff‑layer policy that bounds per‑layer prefetch depth based on empirical profiles and an analytical latency model, guaranteeing just‑in‑time availability without overfetch; and (3) a pipelined runtime with asynchronous prefetch threads and batched I/O to hide loading latency.
\clk{Extensive experiments across datasets, environments, and MoE models demonstrate that \name achieves TPOT speedups of up to 2.04$\times$, 3.10$\times$, and 1.20$\times$ over SD-enabled Mixtral-Offloading, MoE-Infinity, and AdapMoE, respectively. The code is available at https://anonymous.4open.science/r/SpecMoE-2812/.}
\end{abstract}

\section{Introduction}
\label{sec:intro}

\begin{figure}[t]
  \includegraphics[width=1.0\linewidth]{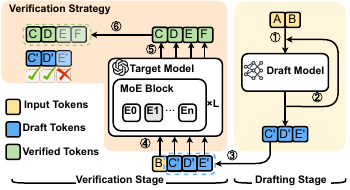}
  \centering
  \caption{MoE-based LLM with speculative decoding }
  \label{fig:spec}
\end{figure}

Large language models (LLMs) such as Llama~\cite{touvron2023llama}, GPT-4~\cite{achiam2023gpt}, and DeepSeek-V2~\cite{liu2024deepseek} have gained widespread popularity in recent years. 
However, their standard decoding process generates tokens one by one in an autoregressive manner~\cite{miao2023towards}, limiting computational parallelism and leading to suboptimal GPU utilization.
To alleviate this bottleneck, recent work has introduced speculative decoding (SD) mechanisms~\cite{chen2023accelerating, gante2023assisted, leviathan2023fast,stern2018blockwise,xia2022speculative}. 
As illustrated in Figure~\ref{fig:spec}, SD decomposes decoding into two stages. In the drafting stage, a smaller and faster model, termed a {\em draft model}, autoregressively generates up to $N$ (usually $N>1$) {\em draft tokens}. In the subsequent verification stage, a {\em target model}, the full LLM used in standard decoding, processes the $N$ draft tokens {\em in parallel}, accepts the longest prefix consistent with its own predictions, and feeds the accepted tokens back to the draft model for the next iteration. 
By reducing the number of sequential target-model decoding steps, SD can significantly accelerate LLM inference.

Meanwhile, the Mixture-of-Experts (MoE) architecture is another optimization technique that replaces dense feed-forward layers with sparse expert networks, activating only a small portion of experts per token. This design significantly reduces computation while maintaining model quality. Combining MoE with SD techniques can further lower the time per output token (TPOT). Existing works~\cite{li2024eagle, svirschevski2024specexec} have demonstrated that SD can significantly accelerate MoE inference. 
However, MoE's sparse structure substantially increases memory requirements, making many MoE models difficult to deploy on a single GPU. For example, Mixtral 8$\times$7B~\cite{jiang2024mixtral} requires approximately 87~GB of memory for inference, far exceeding the 24~GB capacity of a consumer-grade GPU such as the NVIDIA RTX 4090. 

This single-GPU regime is practically important. Although large-scale cloud serving often relies on multi-GPU clusters, many realistic deployments are constrained by cost, privacy, or physical footprint. Examples include departmental or institutional inference services, small-enterprise on-premise deployment, and edge servers close to end users. For these memory-constrained scenarios, expert offloading has become a natural approach to serving MoE models~
\cite{eliseev2023fast,xue2024moe,zhong2024adapmoe,hwang2024pre,kong2023swapmoe}. In this approach, infrequently used experts are stored in CPU memory and loaded into GPU memory on demand. In the same regime, SD is also appealing because it reduces decoding latency without requiring larger accelerators. We focus on single-GPU deployment using pretrained draft models, including quantized variants, without
additional target-specific drafter training. Recent work demonstrates the feasibility of quantized drafting through post-training weight conversion~\cite{georganas2025ml}.

Combining existing expert offloading with SD creates a multi-token verification {\em memory bandwidth bottleneck}. Despite theoretical speedups~\cite{huang2025moesd}, existing methods overlook SD's two-stage structure for prefetching. More draft tokens activate more experts, increasing CPU--GPU transfers (Figure~\ref{fig:activation_rate}). Also, Low acceptance wastes bandwidth on rejected-token experts, hindering accepted-token expert loading.

\noindent{\bf Insights from observations.} Three key observations motivate SD-aware offloading ($\S$\ref{sec: Background and Motivation}). {\em First,} activation analysis shows that neighboring draft tokens tend to activate similar experts, enabling critical-expert prediction (Observation I). {\em Second,} excessive future-layer prefetching without considering I/O contention (e.g., MoE-Infinity, Figure~\ref{fig:pipelining}) reduces prefetching accuracy, increases on-demand loading, and delays computation (Observation II). {\em Third,} non-trivial drafting time permits expert preloading to reduce verification-stage loading latency (Observation III).

We propose \name, a novel SD-aware system accelerating MoE inference through expert prefetching, caching, and computation--communication pipelining. Its core layer-to-layer strategy combines each draft layer's attention output with the corresponding target-layer gate during drafting to predict verification experts, providing additional I/O time for accurate and timely prefetching. However, prefetching all layers risks cache thrashing: deeper-layer experts are consumed later in verification, so prefetching them too early can waste scarce GPU memory and trigger unnecessary evictions before use. We address this through carefully designed system optimizations summarized below as our main contributions.



\begin{itemize}[leftmargin=*]

\item We design drafting-stage speculative prefetching guided by empirical observations and a latency model balancing computation, bandwidth, and memory capacity. \name combines draft-model attention outputs with target-layer gates to predict multi-token verification's critical experts. A profiling-informed cutoff limits prefetching to preceding layers, ensuring just-in-time availability without cache thrashing ($\S$\ref{sec:predictor}).

\item Beyond coarse-grained overlap, our fully pipelined runtime continuously prefetches through an asynchronous worker on a CUDA stream, decoupling transfers from computation. Batched I/O eliminates per-expert synchronization, hiding I/O latency, reducing overhead, and sustaining high bandwidth utilization ($\S$\ref{sec:prefetching}).

\clk{
\item We evaluate \name across three MoE models, four datasets, and three single-GPU environments, reducing TPOT by up to 67.7\% over three SD-enabled MoE offloading baselines.}

\end{itemize}
\section{Background and Motivation}
\label{sec: Background and Motivation}

This section reviews the background on expert offloading in SD-enabled MoE inference and then presents the key empirical observations that directly motivate the design of \name.

\subsection{Expert Offloading in MoE Inference}

\para{MoE-based LLM inference}
uses sparse expert activation to reduce computation, but large expert
weights often require offloading on single-GPU platforms.
Expert loading can dominate inference latency as shown in Figure~\ref{fig:Latency Distribution})~\cite{tang2024hobbit}.
Existing systems mitigate this overhead through prefetching, caching,
and compute--communication overlap~\cite{hwang2024pre,fang2025klotski,
du2024sida,xue2024moe,eliseev2023fast,zhong2024adapmoe}.
However, they do not fully exploit SD's drafting--verification structure
to address expert-transfer bottlenecks during multi-token verification.
This gap motivates the following observations and the design of \name.

\subsection{Key Observations and Design Motivation}
To build an effective prefetching system that alleviates bandwidth bottleneck, two components are essential: (i) an accurate predictor for speculative prefetching and (ii) a pipelined runtime for compute–communication overlap. We base our design on the following observations.

\begin{figure}[t]
    \centering
    \begin{subfigure}{0.45\linewidth}
    \includegraphics[width=\linewidth]{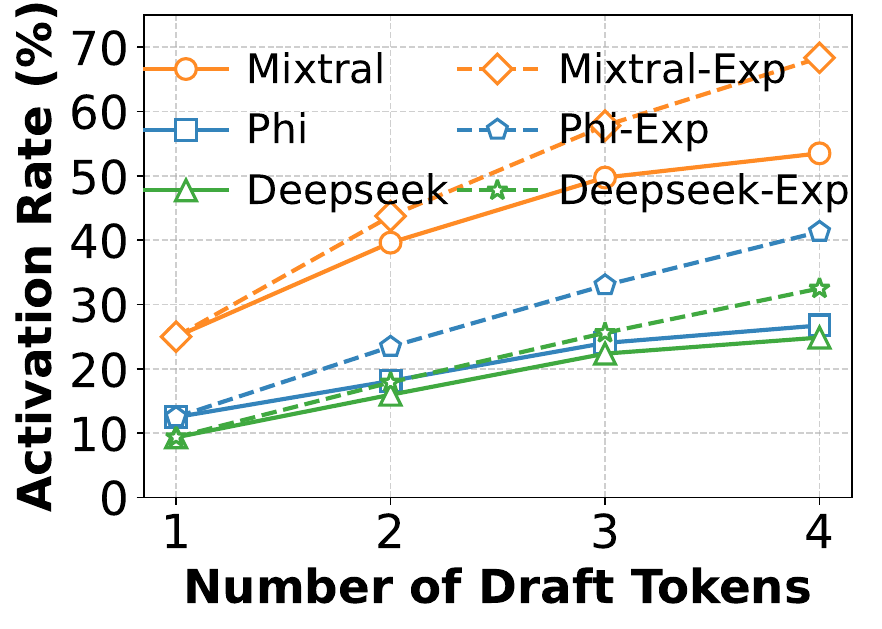}
    \caption{Activation rate of three models and their theoretical expectations (-Exp).}
    \label{fig:activation_rate}
    \end{subfigure}\hfill
    \begin{subfigure}{0.45\linewidth}
    \includegraphics[width=\linewidth]{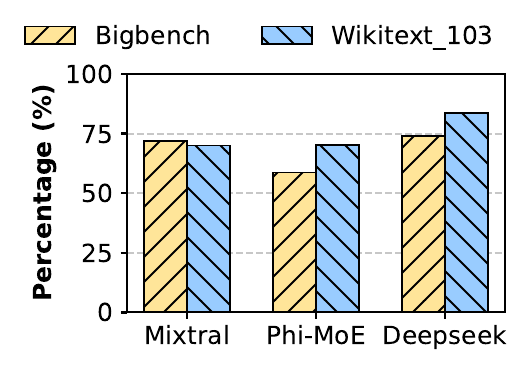}
    \caption{Percentage of token pairs with overlapping expert
activation sets.}
    \label{fig:expert-overlap}
    \end{subfigure}\hfill
    \hspace{0.25cm}
    \caption{Observation I: Neighboring draft tokens share expert activations, enabling prefetching to reduce loading overhead and inference latency.}
    \label{fig:obs2-expert-preditability}
\end{figure}

\para{Observation \uppercase\expandafter{\romannumeral1}: Promising expert preditability for neighboring draft tokens.}
As shown in Figure~\ref{fig:activation_rate}, the
activation rates increase much more slowly than their theoretical expectations under independent and uniformly random activation for each token, as the number of neighboring tokens grows. This indicates that neighboring tokens may activate the same experts.
Figure~\ref{fig:expert-overlap} also shows the high proportions of token pairs with overlapping expert activation sets across two datasets~\cite{ghazal2013bigbench,merity2016pointer} for three models~\cite{jiang2024mixtral,abdin2024phi,liu2024deepseek}. Since the draft model maintains the properties of autoregressive decoding, the generated draft tokens exhibit sequential dependencies, resulting in similar expert activation patterns across neighboring tokens. How to leverage this token-wise similarity to increase expert prediction accuracy becomes the key.

\begin{figure}[t]
    \includegraphics[width=1\linewidth]{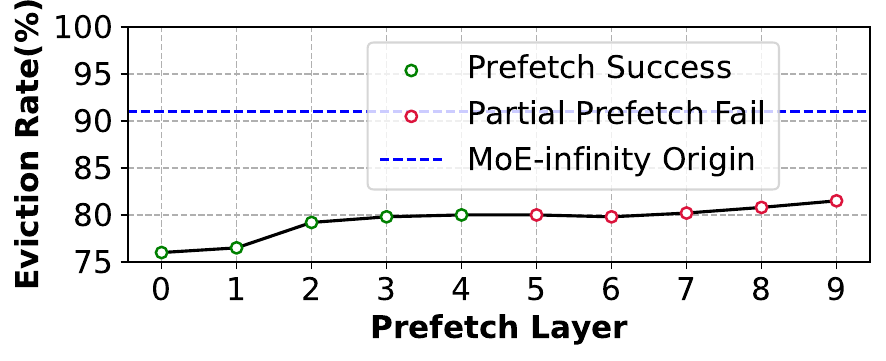}
    \caption{Observation II: Prefetching for lighter layers yields lower eviction rates.}
    \label{fig:Cache Eviction Rate}
\end{figure}

\para{Observation \uppercase\expandafter{\romannumeral2}: Excessive expert prefetching should be avoided.} 
Greedy coarse-grained prefetching~\cite{xue2024moe} generates excessive per-layer tasks, causing bandwidth bottlenecks, blocking on-demand loading, and evicting potentially useful experts under limited GPU memory. To validate this effect, we vary the prefetching depth, i.e., the number of layers allowed for parameter prefetching, which controls the pressure on the prefetch queue, and measures the eviction rate caused by prefetching. Figure~\ref{fig:Cache Eviction Rate} shows worsening GPU cache thrashing. Performance degrades nonlinearly beyond three layers, suggesting a cache-efficiency threshold. When the prefetch depth exceeds six layers, prefetch failures occur due to the memory bandwidth bottleneck, significantly compromising prefetching efficiency. Prefetching should therefore prioritize quality over quantity, accurately identifying and fetching only the most critical experts for the computation of verification.

\begin{figure}[t]
    \centering

    \includegraphics[width=0.95\linewidth]{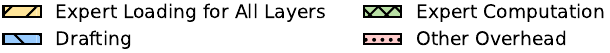}
    \includegraphics[width=0.98\linewidth]{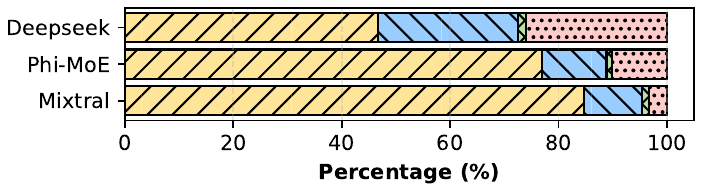}
    \caption{Observation III: Latency distribution of a single decode iteration across three models.}
    \label{fig:Latency Distribution}
\end{figure}

\begin{figure}[t]
  \includegraphics[width=1.0\linewidth]{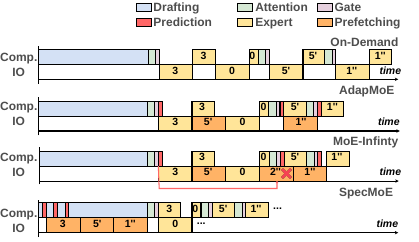}
  \centering
  \caption{Pipeline of four representative mechanisms that offload expert parameters. The number of single quotation indicates the number of layer.}
  \label{fig:pipelining}
\end{figure}

\para{Observation \uppercase\expandafter{\romannumeral3}: Drafting stage provides idle I/O budget for prefetching.}
Expert loading dominates SD-enabled MoE inference latency (69.4\% on average, Figure~\ref{fig:Latency Distribution}), owing to large experts (336, 150, and 16.5~MB in Mixtral 8$\times$7B, Phi-MoE, and Deepseek, respectively) and limited CPU--GPU bandwidth (PCIe 4.0, 32~GB/s). Drafting accounts for another 16.2\% on average while underutilizing I/O bandwidth. Pipeline analysis (Figure~\ref{fig:pipelining}) shows that MoE models often exceed GPU memory, precluding full on-device deployment. Vanilla {\em on-demand expert loading} creates computation-stage bubbles. Gating-based prefetching, such as AdapMoE~\cite{zhong2024adapmoe}, can reduce verification bubbles when extended to SD but leaves drafting-stage bandwidth underutilized. Coarse-grained, history-based prefetching, such as MoE-Infinity~\cite{xue2024moe}, may load incorrect experts, adding overhead. These observations motivate \textbf{drafting-stage prefetching} (\name), which overlaps drafting with expert transfers for the target model's first few layers, improving I/O utilization, reducing pipeline bubbles, and lowering inference latency.

\section{System Design}
\label{sec:predictor and prefetcher}
\subsection{Overview of \name}



\begin{figure}[t!]
  \includegraphics[width=1\linewidth]{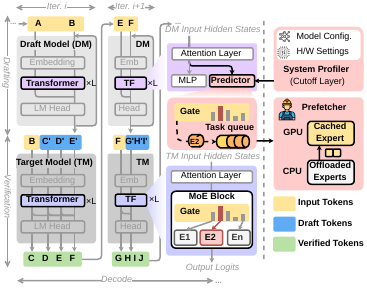}
  \caption{Overview of \name.}
  \label{fig:specMoE-prefetch-overview}
\end{figure}

We propose \name, a novel inference engine for MoE architectures with SD (Figure~\ref{fig:specMoE-prefetch-overview}). It consists of three components: (1) a prediction module that uses a gating predictor to identify critical experts and push them into a prefetching queue (§\ref{sec:predictor}); (2) a profiling module that analyzes expert sizes and hardware to set a cutoff layer, limiting prefetched experts while maximizing overlap with draft computation (§\ref{ssec:cutoff}); and (3) a prefetching module that executes tasks from the queue based on the cutoff layer, overlapping prefetching with draft model computation to deliver the core functionality (§\ref{sec:prefetching}).


\subsection{Cross-Model Expert Predictor}
\label{sec:predictor}
To enable drafting-stage prefetching, we propose an expert predictor that feeds the attention output from each layer of the draft model into the corresponding gating network of target model, which scores each expert and identifies the critical ones for prefetching. 
As shown in Figure~\ref{fig:specMoE-prefetch-overview}, in each layer $l \in [L]$ during drafting, the attention output is fed into our predictor, which directly reuses the gating network of the $l$-th target layer. Since this gating network is well trained, it can be effectively applied in the drafting stage, achieving high accuracy in identifying critical experts while preserving the draft model’s computational flow.

This design is driven by the following two reasons: (1) each drafting stage requires a full forward pass through all layers, mirroring the target model’s inference but occurring earlier; and (2) the attention outputs between draft and target models exhibit high similarity. For example, on WikiText-103~\cite{merity2016pointer}, this similarity remains consistently high across layers, reaching up to 94.59\% for DeepSeek-Lite, 59.82\% for Mixtral 8$\times$7B, and 56.59\% for Phi-3.5-MoE, indicating that the dual-model predictor can provide sufficiently accurate predictions. More detailed measurements and  architecture justifications can be found in Appendix \ref{appdix:similarity-figs}.

\para{Draft-target model configuration.} To maintain SD's efficiency, the architectural compatibility between the draft model and target model is critical. For instance, the optimal draft model corresponding to the target model Mixtral 8x7B~\cite{jiang2024mixtral} would be Mistral 7B~\cite{jung2010mistral}, based on SpecExec~\cite{svirschevski2024specexec}. For target models Phi-3.5-MoE~\cite{abdin2024phi} and Deepseek-Lite~\cite{liu2024deepseek}, we select Phi-mini-MoE and Deepseek-Lite-AWQ as the corresponding draft models. As shown in Table~\ref{tab:model_comparison}, when evaluated on the HumanEval~\cite{li2024humaneval} dataset, the three model pairs achieved the draft token acceptance rate (AC) of 97.42\%, 98.15\%, and 97.01\%, respectively. Such high ACs ensure that most drafted tokens are correct, leading to both high-quality outputs and low inference latency.

\begin{algorithm}[t]
\caption{Expert Prediction and Prefetching Task Queue Management.}
\label{alg:draft_stage}
\SetAlgoVlined
\SetKwInOut{Input}{Input}
\SetKwInOut{Output}{Output}

\Input{Prefetching task queue $Q_{load}$, attention output $s$, cutoff layer $L$, current layer $l$, gates network $Gates$, $k$, critical experts $E_{critical}$, cached queue $Q_{cache}$, $cuda.Event$, $cuda\_expert\_stream$}

\If{MLP of the l-th layer drafting is triggered \& l $\leq$ L}{ \label{line: cutoff layer}

$expert\_scores \gets Gates[l](s)$ \label{line:gating predictor}\;
$E_{critical} \gets TopK\_Index(expert\_scores,\ k)$\; \label{line:topk}
\For{$expert \in E_{critical}$}{\label{line:critical expert}
    \If{$expert \in Q_{cache}$}{
        $E_{critical}.remove(expert)$ \label{line:remove cached expert}\;
    }
}
$cuda.Event.record(cuda\_expert\_stream)$ \label{line:event record}
$Q_{load}.push\_back(E_{critical},\ cuda.Event)$ \label{line:enqueue expert}
}
\end{algorithm}

\subsection{Cutoff Layer Design}
\label{ssec:cutoff}
Prediction and prefetching incur overhead at each layer during drafting, so they should not delay the completion of either drafting generation or target-model verification. To balance prefetching benefits against its cost, we introduce a cutoff layer $L$, meaning that during drafting we prefetch experts only for layers $0$ to $L$ layers based on prediction. This cutoff layer controls the number of prefetched experts and thus the I/O overhead. Our goal is to  minimize the total latency, formulated as $T = T_{drafting} + T_{comp} + T_{I/O}$, where $T_{drafting}$ is the total drafting time, $T_{comp}$ is total verification computation time, and $T_{I/O}$ is the accumulated non-overlapped expert loading time across all verification steps. Moreover, two constraints guide the cutoff: (1) GPU memory, i.e., peak non-expert memory ($M_{peak}$) plus prefetched experts must not exceed capacity ($M_{GPU}$); and (2) time overlap, i.e., expert loading must be fully hidden by computation, otherwise all prefetched tasks up to the cutoff must finish before drafting ends. Formally, for each draft token generation, let $N_{expert} = \sum_{i=0}^{L} k_i$ denote the total number of experts to prefetch. Prefetching must satisfy the following memory and latency constraints:
\begin{align*}
    &M_{peak} + N_{expert} \cdot M_{expert} < M_{GPU},\\
    &\max\{(L-1)t_{l_{comp}} + k_L \cdot t_{E_{I/O}},
    ~N_{expert} \cdot t_{E_{I/O}}\}\\
    &\leq L_{all} \cdot t_{l_{comp}},
\end{align*}
where $k_{i}$ is the number of prefetched experts in layer $i$, $M_{expert}$ is the expert size, $M_{peak}$ is the peak capacity, $M_{GPU}$ denotes the GPU memory volume, $L_{all}$ denotes the total number of transformer blocks in draft model, and $t_{l_{comp}}$ and $t_{E_{I/O}}$ denote the per-layer computation and per-expert loading times, respectively. The last inequality enforces that the chosen cutoff layer allows all expert prefetching to complete during drafting, no matter whether compute time or I/O time is the bottleneck. To stabilize caching, we reserve a fixed number of experts per layer, denoted by $k$ in Algorithm~\ref{alg:draft_stage}. We have $k_i \leq k$, as cached experts are not prefetched. 

Given this formulation, our goal is to solve for $L$ that satisfies these requirements while minimizing the objection function. Because $k_i$ is difficult to predict precisely, we approximate it with $k$ and maximize $L$ under the constraints using profiled $t_{comp}$ and $t_{I/O}$, along with pre-known parameters $M_{expert}$ and $M_{GPU}$. In practice, we pick the top ($k=1$) expert as critical experts for Mixtral 8$\times$7B, the top two ($k=2$) for Phi-3.5-MoE, and the top six ($k=6$) for Deepseek-Lite (line~\ref{line:topk}) based on the following experience. Those values of $k$ align with the number of experts activated per token per layer. Notably, the critical expert count of Mixtral 8$\times$7B does not match its single-token activated expert count. This is because we get high accuracy in critical expert predictions with $k=1$, and the size of one expert in Mixtral 8$\times$7B (336~MB) is far larger than that in other models (150~MB in Phi-MoE and 16~MB in Deepseek), resulting in significantly higher costs when prefetching errors. If $L \leq x$, we set $L = \lfloor x \rfloor$, ensuring all experts up to $L$ are prefetched during drafting to reduce verification latency. Because multiple draft tokens are processed in parallel and often activate overlapping experts, prefetched experts are frequently reused, pushing the system toward optimal performance. 
Overall, the approach reliably reduces end-to-end inference latency in practise.

\para{Interpretation of Algorithm~\ref{alg:draft_stage}.} Integrating the computed cutoff layer $L$ with our cross-model predictor design, \name employs speculative expert prefetching. During drafting, for each layer $l\in \{0, \cdots, L\}$, \name first generates prediction scores for all the experts using the attention outputs $s$ computed in the current layer (line~\ref{line:gating predictor}). It then selects the top-k critical experts based on the gating scores.
For those predicted critical experts that are cached in $Q_{cache}$ within GPU memory, we skip prefetching it (lines \ref{line:critical expert}--\ref{line:remove cached expert}). Otherwise, the rest of the critical experts are then pushed in a prefetching task queue $Q_{load}$ under a CUDA stream. Each enqueue operation  records a synchronization checkpoint $cuda.Event$, which coordinates interaction between the predictor and executor modules (lines \ref{line:event record}, \ref{line:enqueue expert}), as elaborated in $\S$\ref{sec:prefetching}. 

\subsection{Expert Prefetcher}
\label{sec:prefetching}

\begin{figure}[t]
  \centering
  \includegraphics[width=0.95\linewidth]{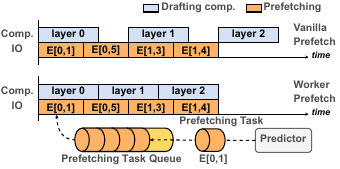}
  \centering
  \caption{Workflow of vanilla-prefetch executor and worker-prefetch executor during the drafting stage.}
  \label{fig:prefetch-executor}
\end{figure}

While prior work~developed accurate prediction modules, it fails to maximize the overlap between expert loading and computation in its prefetching design. 
AdapMoE~\cite{zhong2024adapmoe}, for instance, temporarily stores prefetch information for one expert of the next layer after each prediction step, but delays the actual prefetch until just before the next layer’s experts are required. Since all experts for the current layer are retained in GPU memory, this strategy avoids evicting the prefetched expert. However, prefetching only a single expert offers limited benefit, and prefetching multiple experts per layer triggers I/O synchronization that blocks the subsequent layer’s computation, as shown in Figure~\ref{fig:prefetch-executor}.

\para{Continuous prefetching via worker thread.} If a prefetching mechanism overlaps expert loading only with the computation of the current layer (e.g.,~\cite{zhong2024adapmoe}), it may suffer from stalls: when the computation time of any layer $l$ is insufficient to hide its expert loading time, CUDA memory copy must synchronize and wait for $l$'s expert prefetching before proceeding to the next layer, causing unnecessary overhead (Figure~\ref{fig:prefetch-executor}). To avoid this, we introduce a worker thread, termed Prefetcher, which decouples computation and I/O of the same layer so that they can execute asynchronously, as detailed in Algorithm~\ref{alg:multi_task}. This design enables continuous expert prefetching across the drafting stage, allowing its total computation time to fully cover expert loading.

\begin{algorithm}[t]
\caption{Prefetching Algorithm.}
\label{alg:multi_task}
\SetAlgoVlined
\SetKwInOut{Input}{Input}
\SetKwInOut{Output}{Output}

\Input{Prefetching task queue $Q_{load}$, cache queue $Q_{cache}$, experts to load $E_{load}$, experts to evict $E_{evict}$}


\While{LLM\ inference\ is\ not\  completed}{ \label{line: while do}
    \If{$Q_{load}\ is\ not\ empty$ }{
        //Step 1: fetch the critical expert loading tasks from the queue.\\
        $E_{load},\ cuda.Event \gets Q_{load}.pop()$ \label{line: pop}
        $cuda.Event.wait()$ \label{line: sync}\; 
        $N \gets len(E_{load}) $\;
        //Step 2: select an equal number of evicted experts to replace the prefetched experts.\\
        \For{$i=1$ to $len(Q_{cache})$ }{ \label{line: eviction 1}
            $E_{evict}.append(Q_{cache}[i])$\;
            \If{$len(E_{evict})$ == $N$}{
                break\; \label{line: eviction 2}
            }
        }
        //Step 3: batch-replace the prefetched experts.\\
        $copy\_non\_blocking(E_{load},\ E_{evict})$\;
        \label{line: batch replace}
        \For{$i=1$ to $N$ }{ 
            $Q_{cache}.move\_to\_end(E_{load}[i])$ \label{line: move to end}
        }
    }
}
\end{algorithm}

\para{Queue synchronization for reliable prefetching.} The prefetching and prediction modules jointly maintain a prefetching task queue $Q_{load}$, where each task is associated with a fine-grained synchronization checkpoint $cuda.Event$. An expert loading task is executed by the worker thread only after its information is fully prepared in the queue, avoiding prefetching errors and data loss from incomplete task information. For example, once expert 1 of 0-th layer is selected by the prediction module, as shown in Figure~\ref{fig:prefetch-executor}, its prefetching information is pushed into $Q_{load}$, and a synchronization checkpoint is recorded to confirm successful enqueueing (line~\ref{line:event record} of Algorithm~\ref{alg:draft_stage}). The worker thread (Prefetcher) continuously processes prefetching tasks until inference completes (line~\ref{line: while do} of Algorithm~\ref{alg:multi_task}). Each time, it retrieves expert loading information from $Q_{load}$ and checks synchronization checkpoints $cuda.Event$ for each prefetching task to ensure data integrity (lines \ref{line: pop} of  Algorithm~\ref{alg:multi_task}). 


\para{Batched I/O operations for expert loading.} To further reduce overhead, for the experts to be prefetched in each layer, we batch their I/O operations to run consecutively within the same time period (line~\ref{line: batch replace} of Algorithm~\ref{alg:multi_task}). This minimizes frequent switching between I/O and computation, reducing the kernel overhead of launching the I/O tasks. Meanwhile, an equal number of GPU-cached experts must be batch-selected to replace the prefetched ones, where the cache queue $Q_{cache}$ follows an LRU caching strategy as detailed in $\S$\ref{ssec:cache implementation} (lines \ref{line: eviction 1}--\ref{line: eviction 2}, \ref{line: move to end} of Algorithm~\ref{alg:multi_task}). 

\section{Evaluation} 

\para{Hardware.} Our experiments are performed in three different environments, with configuration shown in Table~\ref{tab: hardware}.

\begin{table}[h]
\renewcommand{\arraystretch}{1.4}
\caption{Hardware environments for evaluation.}
\centering
\resizebox{\linewidth}{!}{
\begin{tabular}{c|cc|cc|cc}
\Xhline{1pt}
\multirow{2}{*}{\textbf{Environment}} & \multicolumn{2}{c|}{\textbf{GPU}} & \multicolumn{2}{c|}{\textbf{CPU}}  & \multicolumn{2}{c}{\textbf{PCIe}} \\ \hhline{~-----}
                                   & \textbf{Model}          & \textbf{Memory}  & \textbf{Model}           & \textbf{Memory} & \textbf{Model}          \\ 
\Xhline{1pt}
\textbf{Env. 1}                      & NVIDIA RTX 3090         & 24 GB  & Intel Xeon Gold 6348  & 512 GB       & \multicolumn{2}{c}{4.0 x 16}           \\ \hline
\textbf{Env. 2}                      & NVIDIA RTX 4090        & 24 GB   & Intel Xeon Gold 6338N  &  1T       & \multicolumn{2}{c}{4.0 x 16}     \\ \hline
\textbf{Env. 3}                      & NVIDIA A100              & 40 GB & Intel Xeon Platinum 8358P  & 1T       & \multicolumn{2}{c}{4.0 x 16}               \\ \hline
\Xhline{1pt}
\end{tabular}
}
\label{tab: hardware}
\end{table}

\begin{table}[t]
    \centering
    \caption{TPOT comparison across different datasets and
    model types in three environments.
    MO indicates Mixtral-Offloading, MI indicates
    MoE-Infinity, AM indicates AdapMoE, SM indicates SpecMoE.}
    \label{tab:tpot_comparison}
    \footnotesize
    \setlength{\tabcolsep}{3pt} 
    \renewcommand{\arraystretch}{1.05}
    \begin{tabular}{@{}ccccccc@{}}
        \toprule
        \textbf{Env.}
        & \textbf{Model Type}
        & \textbf{Dataset}
        & \textbf{MO}
        & \textbf{MI}
        & \textbf{AM}
        & \textbf{SM} \\
        \midrule
        \multirow{6}{*}{1}
        & \multirow{4}{*}{\shortstack{Mixtral\\8$\times$7B}}
        & HumanEval
        & 1.45 & 1.04 & 0.96 & \textbf{0.83} \\
        & & BigBench
        & 1.64 & 1.15 & 1.08 & \textbf{0.93} \\
        & & Wikitext-103
        & 1.60 & 1.13 & 1.05 & \textbf{0.90} \\
        & & MMLU-Pro
        & 1.57 & 1.11 & 1.06 & \textbf{0.89} \\
        \cmidrule(lr){2-7}
        & Phi-3.5-MoE
        & HumanEval
        & 0.50 & 0.74 & 0.34 & \textbf{0.34} \\
        & DeepSeek-Lite
        & HumanEval
        & 0.31 & 0.53 & 0.26 & \textbf{0.24} \\
        \midrule
        \multirow{6}{*}{2}
        & \multirow{4}{*}{\shortstack{Mixtral\\8$\times$7B}}
        & HumanEval
        & 1.42 & 1.11 & 0.91 & \textbf{0.78} \\
        & & BigBench
        & 1.59 & 1.22 & 1.07 & \textbf{0.90} \\
        & & Wikitext-103
        & 1.55 & 1.22 & 1.12 & \textbf{0.94} \\
        & & MMLU-Pro
        & 1.53 & 1.39 & 1.03 & \textbf{0.86} \\
        \cmidrule(lr){2-7}
        & Phi-3.5-MoE
        & HumanEval
        & 0.50 & 0.89 & 0.36 & \textbf{0.30} \\
        & DeepSeek-Lite
        & HumanEval
        & 0.30 & 0.57 & 0.23 & \textbf{0.21} \\
        \midrule
        \multirow{6}{*}{3}
        & \multirow{4}{*}{\shortstack{Mixtral\\8$\times$7B}}
        & HumanEval
        & 1.21 & 0.97 & 0.76 & \textbf{0.68} \\
        & & BigBench
        & 1.36 & 1.08 & 0.84 & \textbf{0.75} \\
        & & Wikitext-103
        & 1.36 & 1.04 & 0.81 & \textbf{0.73} \\
        & & MMLU-Pro
        & 1.47 & 1.05 & 0.81 & \textbf{0.72} \\
        \cmidrule(lr){2-7}
        & Phi-3.5-MoE
        & HumanEval
        & 0.33 & 0.65 & 0.23 & \textbf{0.21} \\
        & DeepSeek-Lite
        & HumanEval
        & 0.24 & 0.36 & 0.20 & \textbf{0.17} \\
        \bottomrule
    \end{tabular}
\end{table}

\para{Models.} We evaluate \name on three target--draft pairs: Mixtral 8$\times$7B~\cite{jiang2024mixtral} (32 transformer blocks with 8-expert MoE layers)--Mistral~\cite{jung2010mistral,svirschevski2024specexec}, Phi-3.5-MoE~\cite{abdin2024phi}--Phi-mini-MoE, and Deepseek-Lite~\cite{liu2024deepseek}--Deepseek-Lite-AWQ. These demonstrate reduced expert-loading latency and expanded drafting-stage prefetching opportunities.

\para{Baselines.}
We compare \name against the following three baselines, where we integrate state-of-the-art MoE offloading methods with SD mechanisms. For fair comparison, we reproduced baselines' Mixtral 8$\times$7B experiments in BF16 while maintaining their original methodologies.

\BULLET \textbf{Mixtral-Offloading+SD}~\cite{eliseev2023fast} adopts offloading with an LRU cache to enable Mixtral 8$\times$7B inference on resource-constrained platforms.

\BULLET \textbf{MoE-Infinity+SD}~\cite{xue2024moe} performs request-level prefetching based on historical sequence-level expert activation patterns.  

\BULLET \textbf{AdapMoE+SD}~\cite{zhong2024adapmoe} employs adaptive gating prediction to prefetch experts for subsequent layers during verification, reducing on-demand loading latency.




\subsection{End-to-End Performance Evaluation} 
\label{ssec:end-to-end}

We evaluate the performance of \name against three advanced baselines to validate its effectiveness and scalability, varying dataset types, model variations, and environments.

\para{Evaluation across environments.}
Across the 18 configurations in Table~\ref{tab:tpot_comparison},
\name reduces TPOT by approximately 40.0\%, 37.6\%, and 12.2\%
on average relative to Mixtral-Offloading+SD, MoE-Infinity+SD,
and AdapMoE+SD, respectively.
\clk{The corresponding peak speedups are approximately 2.04$\times$, 3.10$\times$, and 1.20$\times$. The 2.04$\times$ peak over Mixtral-Offloading+SD occurs with Mixtral 8$\times$7B on MMLU-Pro using A100. The 3.10$\times$ peak over MoE-Infinity+SD occurs with Phi-3.5-MoE on HumanEval using A100. The 1.20$\times$ peak over AdapMoE+SD occurs with Phi-3.5-MoE on HumanEval using RTX~4090. }These are maxima for each baseline across the reported configurations, rather than average speedups. Relative to AdapMoE+SD, the average speedups across the six configurations in each environment are approximately 1.13$\times$, 1.17$\times$, and 1.12$\times$ on RTX~3090, RTX~4090, and A100, respectively.

\para{Evaluation across model types.}
We compare the three models on HumanEval to keep the dataset consistent across model types. For Mixtral 8$\times$7B, TPOT reductions relative to Mixtral-Offloading+SD, MoE-Infinity+SD, and AdapMoE+SD, respectively, are approximately 43.8\%, 29.9\%, and 10.5\% on A100. For Phi-3.5-MoE, \name matches AdapMoE+SD on RTX~3090 at the reported precision, while reducing TPOT by approximately 16.7\% on RTX~4090 and 8.7\% on A100. For DeepSeek-Lite, the corresponding reductions relative to AdapMoE+SD are approximately 7.7\%, 8.7\%, and 15.0\% on RTX~3090, RTX~4090, and A100, respectively. These results show that the magnitude of the improvement depends on both the model and the hardware environment. More detailed evaluations can be found in Appendix~\ref{appdix: all}.

\begin{table}[t]
    \centering
    \caption{Ablation study. Baseline (s) represents the vanilla offloading scheme integrated with SD, with worker prefetch during drafting (wp) and batched I/O operations (b) progressively applied.}
    \label{tab:ablation}

    \small
    \setlength{\tabcolsep}{6pt}
    \renewcommand{\arraystretch}{1.1}

    \begin{tabular}{lccc}
        \toprule
        \textbf{Method}
        & \textbf{Mixtral-8$\times$7B}
        & \textbf{Phi-MoE}
        & \textbf{DeepSeek} \\
        \midrule

        s
        & 1.4094
        & 0.4811
        & 0.3162 \\

        s + wp
        & 0.8383
        & 0.3445
        & 0.1710 \\

        s + wp + b
        & \textbf{0.7804}
        & \textbf{0.3023}
        & \textbf{0.1610} \\

        \bottomrule
    \end{tabular}
\end{table}

\subsection{Ablation Study}
As shown in Figure~\ref{fig:Ablation}, the ablation experiments evaluate the incremental impact of optimization techniques on latency (TPOT) for three models. The baseline, SD-enabled offloading scenario with draft token length = 1, achieves a TPOT of 1.40~s (Mixtral), 0.48~s (Phi-MoE), and 0.32~s (Deepseek). 
Notably, Mixtral 8$\times$7B, Phi-MoE, and Deepseek mitigated on-demand loading overhead through prefetching, achieving 1.68$\times$ , 1.39$\times$, and 1.96$\times$ speedup versus baseline respectively. To implement batched I/O operations, we must build upon the worker prefetch mechanism. As shown in Figure~\ref{fig:Ablation}, our batched I/O approach effectively eliminates the switching overhead between I/O tasks and computation tasks, resulting in speedup ratios of 1.80$\times$, 1.59$\times$, and 1.96$\times$ compared to the baseline Mixtral, Phi-MoE, and Deepseek respectively.


\begin{table}[t]
\renewcommand{\arraystretch}{1.4}
\caption{Hit rate performance on Deepseek across datasets and frameworks. MO indicates Mixtral-Offloading and MI indicates MoE-Infinity.}
\centering
\resizebox{0.45\textwidth}{!}{%
\begin{tabular}{c|cccc}
\Xhline{1pt}
\textbf{Dataset}  & \textbf{MO}    & \textbf{MI}    & \textbf{AdapMoE}    & \textbf{\name} \\
\Xhline{1pt}
\textbf{HumanEval}     & 14.60\% & 16.57\% & 18.74\% & 36.85\% \\ \hline
\textbf{Bigbench}      & 17.50\% & 16.34\% & 21.80\% & 41.25\% \\ \hline
\textbf{Wikitext\_103} & 21.53\% & 17.27\% & 25.15\% & 42.20\% \\ \hline
\textbf{MMLU\_Pro}     & 17.04\% & 16.74\% & 21.72\% & 39.92\% \\ \hline
\textbf{Average}       & 17.67\% & 16.73\% & 21.85\% & 40.06\% \\ \hline
\Xhline{1pt}
\end{tabular}%
}
\label{tab:hit_rate_deepseek}
\end{table}

These results suggest that SpecMoE’s superiority comes from the combined effect of several complementary design choices. First, SpecMoE predicts activated experts with high accuracy, especially for DeepSeek, and dynamically caches critical experts to increase hit rates and reduce costly on-demand loading. Second, SpecMoE maintains a consistent advantage across GPU memory budgets, with the benefit being more noticeable under tighter memory constraints, where accurate prefetching and caching are most helpful. As memory increases, the performance gap narrows because all methods rely less on offloading. Third, unlike AdapMoE, SpecMoE avoids additional synchronization overhead through efficient prediction-based prefetching. Together, these factors explain the performance trends observed above and motivate the more detailed analysis below.

\subsection{Hit Rate Evaluation} 

To validate the effectiveness of \name, we analyze end-to-end latency through the expert hit rate, where a hit is counted when an activated expert is already cached in GPU memory. We report results on three model pairs and four datasets.

We first focus our main discussion on the Deepseek model pair as illustrated in Table~\ref{tab:hit_rate_deepseek}. For Deepseek, \name achieves an average hit rate of 40.06\% (36.85–42.20\%), consistently outperforming all other frameworks across all four datasets. AdapMoE ranks second with an average hit rate of 21.85\%, followed by Mixtral-Offloading (17.67\%) and MoE-Infinity (16.73\%). This confirms that \name’s prediction module attains high accuracy on Deepseek: with six experts cached per layer, \name can accurately prefetch critical experts during drafting without introducing additional overhead, thereby yielding substantially higher hit rates. 
\section{Discussion and Conclusion}
\para{Discussion: copy-back overhead.}
Mixtral-Offloading~\cite{eliseev2023fast} partitions experts between GPU and CPU, requiring copy-back upon GPU eviction. Following AdapMoE~\cite{zhong2024adapmoe}, we instead retain CPU copies of all experts, eliminating copy-back at the cost of additional host memory.

\para{Conclusion.}
This paper presents \name, an SD-enabled MoE-based LLM inference system for resource-constrained environments.
\name exploits the two-stage structure of speculative decoding for accurate draft-stage prefetching, and introduces a pipelined runtime with asynchronous prefetching and batched I/O to mitigate bandwidth bottlenecks.
\clk{Extensive experiments across diverse datasets, environments, and MoE-based models show that \name achieves TPOT speedups of up to 2.04$\times$, 3.10$\times$, and 1.20$\times$ over SD-enabled Mixtral-Offloading, MoE-Infinity, and AdapMoE, respectively.} 
%


\section{Limitations}
\name targets memory-constrained, single-GPU MoE inference. Our evaluation covers three model pairs and three hardware environments. Its effectiveness on other architectures, ulti-GPU deployments, and high-concurrency workloads remains to be established.

The cross-model expert predictor relies on compatible draft--target layer structures and sufficiently aligned hidden representations. Its effectiveness may decrease for less aligned model pairs, while structurally different drafters, such as EAGLE-style models, require adapting the predictor.

Cutoff-layer selection uses offline profiling and an approximate latency model. Changes in available memory, transfer bandwidth, or workload may require recalibration. The resulting cutoff is therefore a practical heuristic rather than a guarantee of optimal scheduling under all runtime conditions.

Finally, our experiments primarily evaluate inference latency. They do not comprehensively characterize throughput under concurrent requests, energy consumption, or output quality across tasks. The observed speedups should thus be interpreted within the evaluated deployment settings.
\bibliography{custom}
\section{Appendix}
\label{appdix: all}

\subsection{Appendix on implementation} 
\label{sec:implement}
We implement a prototype of \name using PyTorch and Transformer-based components from the Hugging Face library~\cite{wolf-etal-2020-transformers}. Systems such as vLLM~\cite{kwon2023efficient} and FlexGen~\cite{sheng2023flexgen} optimize KV-cache management and general memory offloading, rather than MoE-specific expert offloading. Although they are complementary to our design, they are not direct baselines for our setting. We therefore leave their integration with \name to future work and restrict our evaluation to systems that explicitly target expert offloading.

\subsubsection{Prediction Module Integration}
Recall that the draft model first generates $N$ draft tokens per iteration, which are then verified in parallel by the target model. To integrate our drafting-stage predictor in to this workflow without modifying the original drafting execution flow, we insert hook functions into each layer of the draft model. These hooks capture attention outputs during execution and feed them into the target model's gating network, which generates scores for us to identify critical experts for prefetching. Moreover, our measurements show that the predictor adds negligible overhead (0.39\% on average).

\subsubsection{SD Implementation and Configuration} 
SD is particularly well-suited for low-batchsize inference scenarios, especially when optimizing end-to-end latency for individual user requests. Existing approaches such as Medusa~\cite{cai2024medusa} and SpecExec~\cite{svirschevski2024specexec} consistently set the batch size to 1, to maximize the optimization of the inference latency. Eagle~\cite{li2024eagle}, a state-of-the-art SD framework, demonstrates the most significant acceleration at batch size = 1 in its batch-scaling experiment. In line with these findings, \name also adopts batch size = 1 to demonstrate its maximal optimization of inference efficiency for SD-enabled and MoE-based LLMs. For the same reason, we do not include these SD methods as direct baselines: they are designed to accelerate dense-model decoding itself, whereas our focus is orthogonal, i.e.g, reducing expert-loading overhead in offloaded MoE inference. We also leave a systematic study of larger batch sizes, where batching and scheduling effects become entangled with offloading behavior, to future work.

\subsubsection{Computation Implementation} 
To demonstrate the effectiveness of prefetching, \name modifies the conventional expert computation order in Transformers~\cite{wolf-etal-2020-transformers}, prioritizing the GPU-cached experts for computation. Taking Mixtral 8$\times$7B as an example, traditional implementation~\cite{wolf-etal-2020-transformers} processes tokens sequentially from Expert 0 to Expert 7. However, to reduce GPU memory switching overhead, \name adopts the existing approaches like ProMoE~\cite{song2024promoe} and Klotski~\cite{fang2025klotski}, employing the reordered computation: experts already cached in GPU memory are prioritized for immediate token processing, while the remaining experts cached in CPU memory are loaded on demand into GPU memory for subsequent computations.

\subsubsection{Cache Implementation} 
\label{ssec:cache implementation}
\name employs the LRU caching strategy to optimize expert eviction in MoE layers. We leverage a cache queue $Q_{cache}$ to track the expert access order. For new experts not in the queue, we evict the head-of-queue expert, replace it with the new one, and push it to the tail. For existing experts in the queue, we first remove them and then reinsert them from the back. Based on the LRU caching strategy, this mechanism prevents recently prefetched experts from being evicted.


\subsubsection{Extension of Distributed Deployment.} 
In practice, MoE-based LLMs may also be deployed on multiple GPUs within a single node via expert parallelism. Even in this setting, expert offloading to CPU memory may still be necessary due to the large number of experts and their memory footprint. Compared with the single-GPU setting, multi-GPU deployment introduces additional GPU--GPU communication. Since high-bandwidth interconnects such as NVLink/NVSwitch provide much higher bandwidth than CPU--GPU transfers, GPU--CPU data movement remains the dominant bottleneck in our target setting, and our current cutoff-layer formulation therefore focuses on GPU-CPU transfer latency. To support multi-GPU execution, the latency model can be extended with an additional GPU-GPU transfer term when a prefetched expert resides on a different GPU from the one executing the current computation, while the memory constraint is generalized to per-GPU cache capacities under a shared GPU memory pool. Accordingly, \name can maintain a global memory manager across GPUs, attach GPU identifiers to prefetched experts, and jointly determine their prefetch timing and placement, while applying the same GPU-aware LRU policy for retention, eviction, and migration. Thus, the core cutoff-layer and cache-management logic of \name can be naturally extended to multi-GPU execution.

\subsubsection{Generality of Cross-model Design}
Our structure-consistent and cross-model-aligned design is not restricted to draft and target models with identical Transformer depths. In practice, draft-target pairing is mainly guided by two metrics: (1) draft token acceptance rate and (2) draft model size, which determines drafting-stage latency. While identical architectures offer ideal one-to-one alignment, practical deployments often yield different layer counts (typically $m < n$, where $m$ and $n$ denote the numbers of layers in the draft and target models, respectively). To handle this case, we perform offline profiling to build deterministic mappings from draft-layer  attention outputs to corresponding target gates, and then use the target model’s gating mechanism for expert prediction. As a result, one draft-layer's attention output to inform expert activation in multiple subsequent target layers, naturally extending our design to heterogeneous draft–target configurations.

\subsection{Appendix on Draft-Target Similarity}
\label{appdix:similarity-figs}
\begin{figure}[t]
    \centering
    \begin{subfigure}{1\linewidth}
    \centering
    \includegraphics[width=0.8\linewidth]{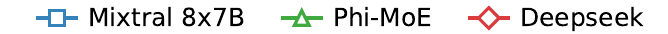}
    \label{fig:sim-legend}
    \end{subfigure}\hfill
  \begin{subfigure}[b]{1\linewidth}
    \includegraphics[width=\linewidth]{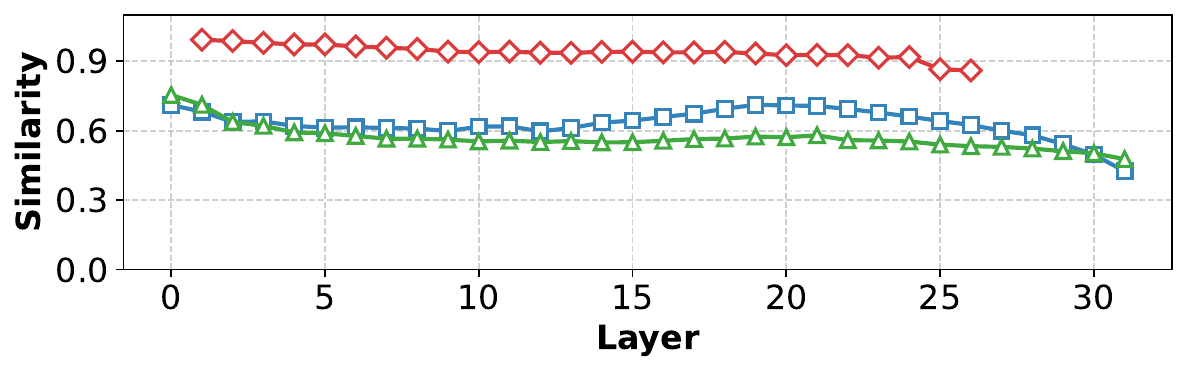}
    \caption{Cosine similarity.}
    \label{fig:cosine_similarity}
  \end{subfigure}
\begin{subfigure}[b]{1\linewidth}
    \includegraphics[width=\linewidth]{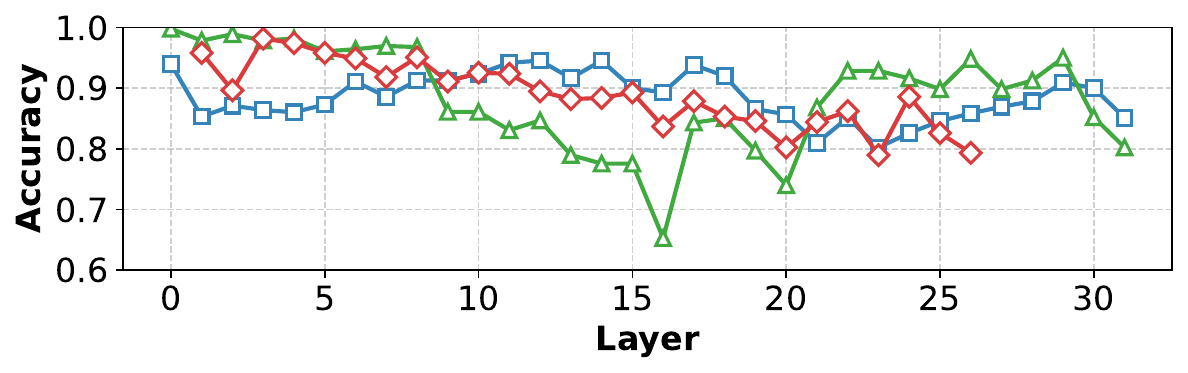}
    \caption{Prediction accuracy.}
    \label{fig:prediction_accuracy}
  \end{subfigure}
  \caption{Cosine similarity and prediction accuracy across layers between draft and target models.}
  \label{fig:layer_analysis}
\end{figure}

Figure~\ref{fig:cosine_similarity} shows consistently high similarity across layers on WikiText-103~\cite{merity2016pointer}, indicating sufficiently accurate prediction using the dual-model-based predictor. Figure~\ref{fig:prediction_accuracy} confirms this, with DeepSeek-Lite achieving 88.94\% top-1 expert prediction accuracy on average across layers, and Mixtral and Phi-3.5-MoE maintaining approximately 88\% accuracy. Such similarity between drafting and verification might stem from the architecture similarity.

\para{Draft-target model configuration.} To maintain SD's efficiency, the architectural compatibility between the draft model and target model is critical. For instance, the optimal draft model corresponding to the target model Mixtral 8x7B~\cite{jiang2024mixtral} would be Mistral 7B~\cite{jung2010mistral}, based on SpecExec~\cite{svirschevski2024specexec}. For target models Phi-3.5-MoE~\cite{abdin2024phi} and Deepseek-Lite~\cite{liu2024deepseek}, we select Phi-mini-MoE and Deepseek-Lite-AWQ as the corresponding draft models. As shown in Table~\ref{tab:model_comparison}, when evaluated on the HumanEval~\cite{li2024humaneval} dataset, the three model pairs achieved the draft token acceptance rate (AC) of 97.42\%, 98.15\%, and 97.01\%, respectively. Such high ACs ensure that most drafted tokens are correct, leading to both high-quality final outputs and low inference latency. 

As shown in Table~\ref{tab:model_comparison}, the two models share similar architectures (experts per layer, hidden size, and number of layers), making a cross-model, layer-to-layer mapping feasible.


\begin{table}[h]
\centering
\caption{Draft and target model architectures. Para. = model parameter size; \#E/L = number of experts per layer; FFN = expert feed-forward size; Hid. = hidden size; \#L = number of transformer blocks; AC = acceptance rate with SD.}
\label{tab:model_comparison}
\resizebox{\linewidth}{!}{
\begin{tabular}{l|cccccc} 
\Xhline{1pt}
\textbf{Model} & \textbf{Para.} & \textbf{\#E/L} & \textbf{FFN} & \textbf{Hid.} & \textbf{\#L} & \textbf{AC}\\ \hhline{~}
\Xhline{1pt}
\textbf{Mistral 7B} & 7B  & N/A & 14336 & 4096 & 32 & \multirow{2}{*}{97.42\%} \\
\textbf{Mixtral 8x7B} & 45B & 8   & 14336 & 4096 & 32 &  \\ \hline
\textbf{Phi-mini-MoE} & 8B  & 16 & 960 & 4096 & 32 & \multirow{2}{*}{98.15\%} \\
\textbf{Phi-3.5-MoE} & 42B & 16   & 6400 & 4096 & 32 & \\
\hline
\textbf{Deepseek-Lite-AWQ} & 3B  & 64 & 1408 & 2048 & 27 & \multirow{2}{*}{97.01\%} \\
\textbf{Deepseek-Lite} & 15B & 64   & 1408 & 2048 & 27 &  \\

\Xhline{1pt}
\end{tabular}
}
\end{table}

\begin{table*}[t]
\renewcommand{\arraystretch}{1.4}
\caption{Hit rate performance across datasets, models, and frameworks. MO indicates Mixtral-Offloading and MI indicates MoE-Infinity.}
\centering
\resizebox{\textwidth}{!}{
\begin{tabular}{c|cccc|cccc|cccc}
\Xhline{1pt}
\multirow{2}{*}{\textbf{Dataset}}  & \multicolumn{4}{c|}{\textbf{Mixtral 8$\times$7B}} & \multicolumn{4}{c|}{\textbf{Phi-MoE}} & \multicolumn{4}{c}{\textbf{Deepseek}} \\ \hhline{~------------}
                                   & \textbf{MO}    & \textbf{MI}    & \textbf{AdapMoE}    & \textbf{\name} & \textbf{MO}    & \textbf{MI}    & \textbf{AdapMoE}    & \textbf{\name} & \textbf{MO}    & \textbf{MI}    & \textbf{AdapMoE}    & \textbf{\name}  \\ 
\Xhline{1pt}
\textbf{HumanEval}                & 15.08\%        & 16.01\%        & 41.83\%       & 18.93\%     & 35.37\%        & 15.28\%        & 56.22\%        & 44.31\%     & 14.60\%        & 16.57\%        & 18.74\%        & 36.85\%    \\ \hline
\textbf{Bigbench}                 & 15.14\%        & 15.83\%        & 42.82\%        & 21.39\%     & 22.36\%        & 14.71\%        & 45.14\%        & 41.38\%     & 17.50\%        & 16.34\%        & 21.80\%        & 41.25\%  \\ \hline
\textbf{Wikitext\_103}            & 15.14\%        & 15.76\%        & 42.55\%        & 21.19\%     & 28.37\%        & 14.62\%        & 50.39\%       & 43.22\%     & 21.53\%        & 17.27\%        & 25.15\%        & 42.20\% \\ \hline
\textbf{MMLU\_Pro}                & 14.73\%        & 15.87\%        & 41.35\%        & 21.06\%     & 24.30\%        & 14.07\%        & 45.97\%        & 41.97\%    & 17.04\%        & 16.74\%        & 21.72\%        & 39.92\% \\ \hline
\textbf{Average}                & 15.02\%        & 15.87\%        & 42.14\%        & 20.89\%       & 27.60\%        & 14.67\%        & 49.43\%       & 42.72\%     & 17.67\%        & 16.73\%        & 21.85\%        & 40.06\%\\ \hline
\Xhline{1pt}
\end{tabular}
}
\label{tab:hit_rate}
\end{table*}
\begin{figure}[t]
   \centering
  \includegraphics[width=0.5\linewidth]{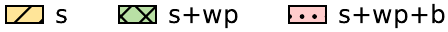}
  \label{fig:Ablation-legend}
  \includegraphics[width=1\linewidth]{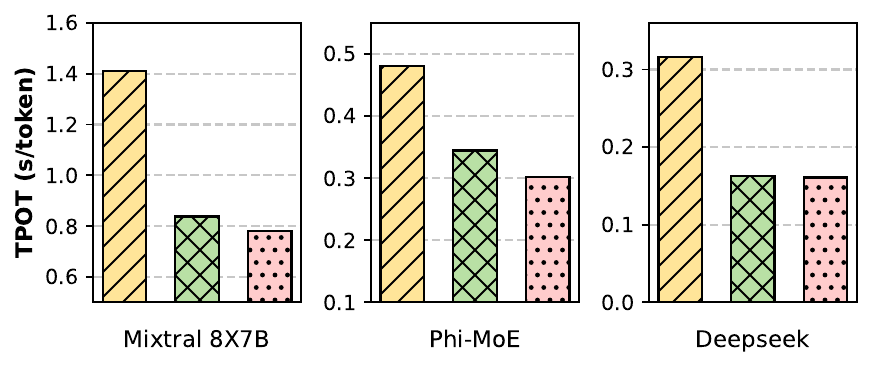}
  \centering
  \caption{Ablation study. Baseline (s) represents the vanilla offloading scheme integrated with SD, worker prefetch during drafting (wp) and batched I/O operations (b) applied.}
  \label{fig:Ablation}
\end{figure}

\subsection{Ablation Study}
As shown in Figure~\ref{fig:Ablation}, the ablation experiments evaluate the incremental impact of optimization techniques on latency (TPOT) for three models. The baseline, SD-enabled offloading scenario with draft token length = 1, achieves a TPOT of 1.40~s (Mixtral), 0.48~s (Phi-MoE), and 0.32~s (Deepseek). 
Notably, Mixtral 8$\times$7B, Phi-MoE, and Deepseek mitigated on-demand loading overhead through prefetching, achieving 1.68$\times$ , 1.39$\times$, and 1.96$\times$ speedup versus baseline respectively. To implement batched I/O operations, we must build upon the worker prefetch mechanism. As shown in Figure~\ref{fig:Ablation}, our batched I/O approach effectively eliminates the switching overhead between I/O tasks and computation tasks, resulting in speedup ratios of 1.80$\times$, 1.59$\times$, and 1.96$\times$ compared to the baseline Mixtral, Phi-MoE, and Deepseek respectively.


\subsection{Hit Rate Evaluation} 

To validate the effectiveness of \name, we analyze end-to-end latency through the expert hit rate, where a hit is counted when an activated expert is already cached in GPU memory. We report results on three model pairs and four datasets.

Besides Deepseek, we further report hit rate across all two model pairs: Mixtral 8$\times$7B and Phi-3.5-MoE under four frameworks: Mixtral-Offloading (MO), MoE-Infinity (MI), AdapMoE, and \name, as summarized in Tables~\ref{tab:hit_rate}. MO originally only supported a quantized Mixtral 8$\times$7B model~\cite{dvmazur}. We extend its source code by adjusting weight configurations and optimizing its offloading logic to support Phi-3.5-MoE~\cite{abdin2024phi}, and DeepSeek-Lite~\cite{liu2024deepseek}. We similarly extend AdapMoE and MoE-Infinity, which originally had limited model compatibility, to cover the same three models. AdapMoE consistently achieves higher hit rate than Mixtral-Offloading and MoE-Infinity, and even exceeds \name on both Mixtral 8$\times$7B and Phi-MoE. Nevertheless, its end-to-end latency remains higher than \name, indicating the overall performance not only depends on hit rate but also on how prefetching is scheduled and overlapped with computation: AdapMoE's prefetching incurs extra synchronization and prefetching overhead, whereas \name reduces this cost through drafting-stage prefetching, asynchronous execution, and batched I/O, achieving a better balance between prediction accuracy and prefetch efficiency.

\begin{figure}[t]
    \centering
    \begin{subfigure}{1\linewidth}
    \centering
    \includegraphics[width=1\linewidth]{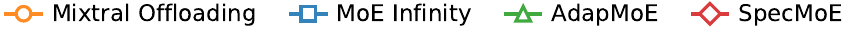}
    \label{fig:TPOT-legend}
    \end{subfigure}\hfill
    \begin{subfigure}{0.95\linewidth}
    \includegraphics[width=\linewidth]{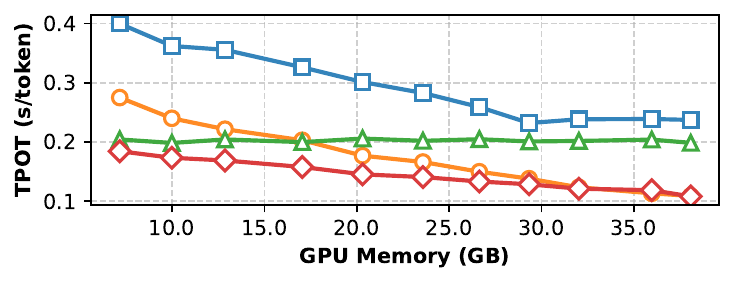}
    \end{subfigure}\hfill
    \caption{TPOT vs. GPU memory for the DeepSeek draft-target model pair on dataset HumanEval in Env. 3.}
    \label{fig:memory}
\end{figure}

\subsection{Impact of GPU Memory Budgets} 

In this section, we evaluate the performance of \name and other baselines varying GPU memory budgets. As GPU memory increases from 7~GB to 39~GB shown in Figure~\ref{fig:memory}, Mixtral-Offloading steadily reduces TPOT from 0.27~s to 0.10~s. MoE-Infinity improves slightly, lowering latency from 0.39~s to 0.28~s at 24 GB, but stabilizes around 0.23~s for higher memory. AdapMoE maintains stable latency, ranging from 0.19~s to 0.20~s across all memory levels, with minimal gains from additional resources. In contrast, \name achieves the lowest or comparable TPOT across the evaluated memory budgets, with larger gains under tighter memory constraints.

These results suggest that SpecMoE’s superiority comes from the combined effect of several complementary design choices. First, SpecMoE predicts activated experts with high accuracy, especially for DeepSeek, and dynamically caches critical experts to increase hit rates and reduce costly on-demand loading. Second, SpecMoE maintains a consistent advantage across GPU memory budgets, with the benefit being more noticeable under tighter memory constraints, where accurate prefetching and caching are most helpful; as memory increases, the performance gap narrows because all methods rely less on offloading. Third, unlike AdapMoE, SpecMoE avoids additional synchronization overhead through efficient prediction-based prefetching. Together, these factors explain the performance trends observed above and motivate the more detailed analysis below.

\begin{figure}[t]
   \centering
  \includegraphics[width=0.97\linewidth]{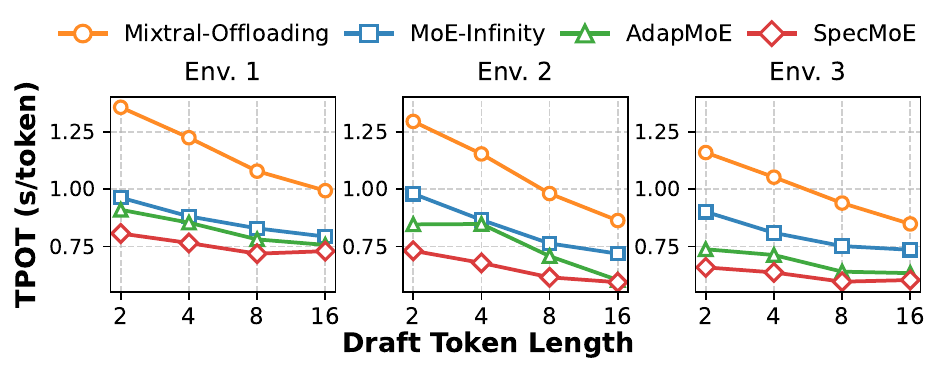}
  \centering
  \caption{TPOT comparison varying draft token length per drafting stage: SpecMoE consistently achieves the lowest TPOT across three environments, though performance gaps narrow slightly with longer draft token length.}
  \label{fig:TPOT comparison varying draft token length per drafting stage}
\end{figure}

\subsection{Impact of the Draft Token Length} 
To validate the effectiveness of \name using SD, we vary the draft token length and compare the performance of four framework across three environments using the HumanEval~\cite{li2024humaneval} dataset on Mixtral 8$\times$7B.

As shown in Figure~\ref{fig:TPOT comparison varying draft token length per drafting stage}, \name consistently outperforms all baselines across RTX 3090, RTX 4090, and A100 environments, achieving the lowest TPOT in all test cases. Its advantage is most pronounced on 3090, where it delivers 15-20\% lower TPOT than AdapMoE, the top-performing baseline, while maintaining a 5-8\% lead even on high-performance A100 systems. 
\name shows stronger scalability with longer draft tokens and consistent hardware optimization. As draft token length increases, the performance gap with other baselines narrows naturally due to SD. Longer draft token length requires activating more experts, potentially reaching full-expert activation, which increases expert loading overhead. In such case, expert prediction accuracy becomes less critical, shifting focus to overlapping loading with computation. This constrains SpecMoE's optimization space, yielding only marginal gains. However, longer draft token length reduces target model iterations while providing prefetching time during drafting to hide expert loading time, ultimately maintaining \name's performance advantage.


\begin{figure}[t]
   \centering
  \includegraphics[width=1\linewidth]{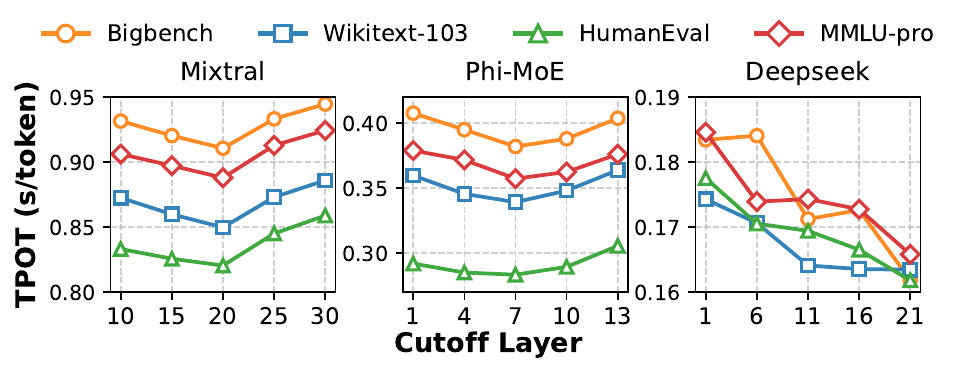}
  \centering
  \caption{TPOT comparison varying the value of cutoff layer: a U-shaped cutoff layer-TPOT relationship for Mixtral/Phi-MoE, but linear latency reduction for DeepSeek with increasing cutoff layers.}
  \label{fig: cutoff layer}
\end{figure}

\subsection{Impact of the Cutoff Layer} 

To validate that prefetched excessive experts impacts end-to-end performance, we adjusted cutoff layer, as illustrated in Figure~\ref{fig: cutoff layer}.

For Mixtral 8$\times$7B, The experiments across all benchmarks demonstrate a distinct U-shaped relationship with increasing cutoff layer, which indicates that the end-to-end performance is increasing as cutoff layer rises from 0 to around 20. Then, it is followed by a recovery trend beyond at the point of 20 where higher prefetch counts ultimately lead to worse TPOT across different datasets. This suggests that excessive prefetching results in poor performance. This trend highlights the importance of balancing the cutoff layer to maximize the benefit of \name. When we switch to the Phi-MoE model, the relationship between the cuoff layer and end-to-end performance also exhibits this U-shaped pattern. It is resulted from that the drafting stage duration is unable to fully overlap with excessive expert loading times, mirroring the fundamental limitation we observed in the architecture of Mixtral 8$\times$7B.
However, when testing with the DeepSeek model, Figure~\ref{fig: cutoff layer} shows a consistently improving trend as cutoff layer increases, unlike the U-shaped pattern observed earlier. This linear progression occurs because the drafting stage duration fully overlaps with the expert loading time of each layer, avoiding the synchronization overhead between the drafting stage and the target stage, which would arise from excessive prefetching. Therefore, progressively increasing the cutoff layer achieves sustained latency reduction without considering the performance optimization threshold observed in other models, demonstrating that the Deepseek model pair is able to capitalize on the aggressive prefetching strategy.

Although cutoff-layer selection is model-dependent, it can be readily determined using profiling tools such as NVIDIA Nsight. By combining offline profiling with theoretical modeling, we can quickly identify the optimal cutoff layer for a given model based on its computation and communication latency. More importantly, \S\ref{ssec:end-to-end} shows that our method is data-independent, demonstrating robustness to input distribution variations.

\subsection{Lessons and Discussion} 
\para{Large batch sizes amplify I/O bottlenecks.}
Eagle-3~\cite{li2025eagle} shows that consumer-grade frameworks like vLLM~\cite{kwon2023efficient} and SGLang~\cite{zheng2024sglang} achieve higher throughput than Eagle~\cite{li2024eagle} at large batch sizes. For MoE-based LLMs, however, larger batches trigger more on-demand loading of activated experts, making I/O overhead more dominant and reducing the benefit of batching, especially because requests in the same batch typically activate uncorrelated expert sets, which undermines expert reuse and caching.

\para{Sequential drafting simplifies analysis but limits generality.}
Speculative decoding can generate draft tokens sequentially~\cite{leviathan2023fast} or as trees~\cite{li2024eagle,miao2024specinfer}. \name currently adopts greedy decoding to maintain system stability and a high verification acceptance rate. As future work, we plan to explore sampling-based decoding and its effect on expert activation, with a focus on temperature: lower temperatures bias toward high-probability drafts that are more likely to be accepted, whereas higher temperatures admit more low-probability drafts, increasing rejection, wasting draft computation, and complicating expert activation prediction.

\end{document}